\documentclass[sigplan,nonacm]{acmart}

\setcopyright{none}
\renewcommand\footnotetextcopyrightpermission[1]{}
\usepackage{makecell}
\usepackage{algorithm}
\usepackage{algpseudocode}
\usepackage{subcaption}
\AtBeginDocument{%
  }

\begin{document}

\title{ReMP: Low-Downtime Runtime Model-Parallelism Reconfiguration for LLM Serving}

\author{Haipeng Yuan}
\email{yuanhaipeng24z@ict.ac.cn}
\affiliation{%
  \institution{Institute of Computing Technology, Chinese Academy of Sciences}
  \city{Beijing}
  \country{China}
}

\author{Kaining Zheng}
\email{zkn0823@163.com}
\affiliation{%
  \institution{Institute of Computing Technology, Chinese Academy of Sciences}
  \city{Beijing}
  \country{China}
}

\author{Yongshu Bai}
\email{ybai@zhejianglab.org}
\affiliation{%
  \institution{Zhejiang Lab}
  \city{Hangzhou}
  \country{China}
}

\author{Yuchen Zhang}
\email{zhangyuchen24s@ict.ac.cn}
\affiliation{%
  \institution{Institute of Computing Technology, Chinese Academy of Sciences}
  \city{Beijing}
  \country{China}
}

\author{Yunquan Zhang}
\email{zyq@ict.ac.cn}
\affiliation{%
  \institution{Institute of Computing Technology, Chinese Academy of Sciences}
  \city{Beijing}
  \country{China}
}

\author{Baodong Wu}
\email{wubd.cs@gmail.com}
\affiliation{%
  \institution{Infinigence AI}
  \city{Beijing}
  \country{China}
}

\author{Xiang Gao}
\email{gaoxiang@zhejianglab.org}
\affiliation{%
  \institution{Zhejiang Lab}
  \city{Hangzhou}
  \country{China}
}

\author{Daning Cheng}
\email{553003893@qq.com}
\affiliation{%
  \institution{Institute of Computing Technology, Chinese Academy of Sciences}
  \city{Beijing}
  \country{China}
}

\renewcommand{\shortauthors}{Yuan et al.}

\begin{abstract}
Current large language model (LLM) inference systems universally deploy ultra-large-scale models using a combination of Tensor Parallelism (TP) and Pipeline Parallelism (PP). However, existing systems treat the model parallelism topology as a static configuration that cannot be flexibly adjusted at runtime. This rigid design creates a fundamental contradiction with the dynamically changing inference workloads in real-world scenarios. State-of-the-art systems lack online reconfiguration capabilities and can only switch configurations by restarting the service, resulting in tens of seconds of service interruption, KV cache loss, and prohibitive recomputation overhead. To address this problem, this paper presents ReMP, a runtime model parallelism reconfiguration framework that supports low downtime. ReMP achieves dynamic adjustment through three key techniques: (1) decoupling the model parallelism topology from runtime state to avoid full service reconstruction; (2) designing a two-dimensional KV cache migration mechanism to preserve reusable cache states after TP/PP changes; and (3) implementing end-to-end online reconfiguration. Experiments demonstrate that ReMP completes the tested topology switches in 1.0--7.4 seconds on models ranging from 7B to 70B parameters. On H100, ReMP is 2.5--25.8$\times$ faster than restart. Moreover, ReMP significantly outperforms fixed configurations under dynamic workloads, delivering superior performance in terms of TTFT, TPOT, and output throughput.

\end{abstract}



\maketitle

\section{Introduction}

\begin{figure}[htbp]
    \centering
    \begin{subfigure}{0.22\textwidth}
        \centering
        \includegraphics[width=\linewidth]{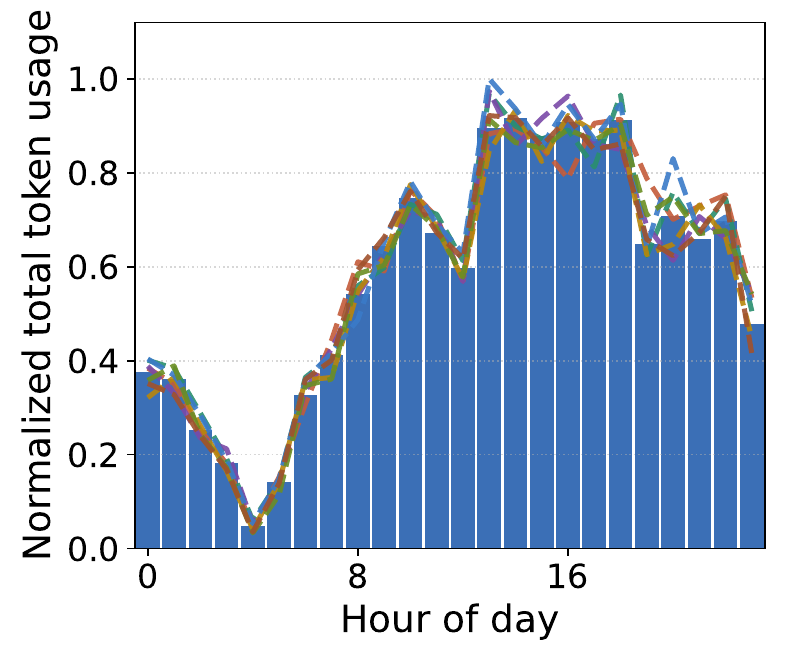}
        \caption{Token usage profile of Provider~A.}
        \label{fig:token-usage-a}
    \end{subfigure}
    \begin{subfigure}{0.22\textwidth}
        \centering
        \includegraphics[width=\linewidth]{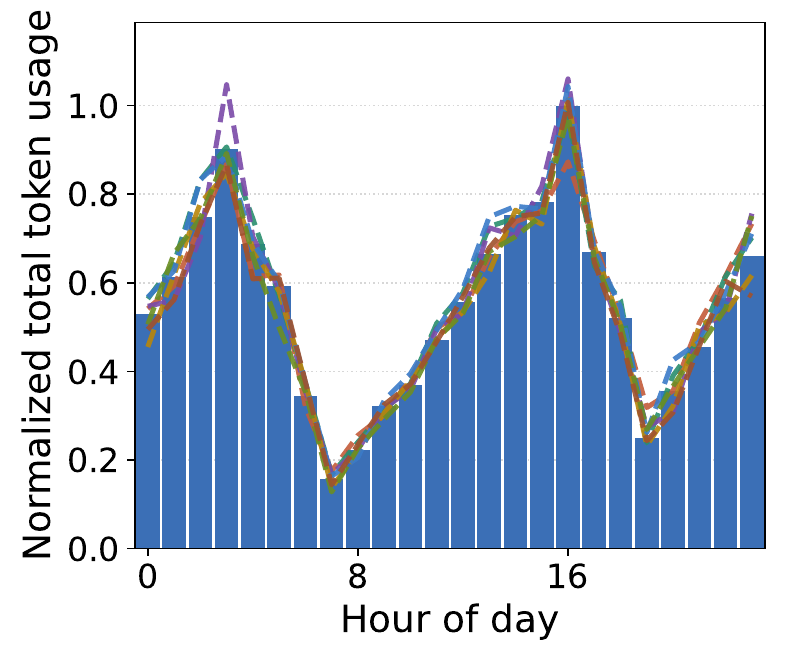}
        \caption{Token usage profile of Provider~B.}
        \label{fig:token-usage-b}
    \end{subfigure}
    \caption{Token usage of one model on two production LLM services, measured across time periods within the same day. Blue bars denote the average traffic over all observed days, and colored dashed lines denote the real traffic of individual days. Traffic is normalized by setting the value at 10:00 to 1, with other time points scaled proportionally.}
    \label{fig:token-usage}
    \Description{Two bar-and-line charts showing normalized token usage over the hours of a day for two production large language model services. In both services, traffic varies substantially between time periods, and the per-day dashed lines follow a similar daily shape.}
\end{figure}

In recent years, Large Language Models (LLMs) have advanced rapidly. However, real-world LLM serving workloads exhibit high diversity and dynamism. On one hand, different application scenarios entail distinct performance objectives, shown in Table~\ref{tab:workload}; on the other hand, request volumes and token usage in production environments fluctuate over time, as illustrated in Figure~\ref{fig:token-usage}. Consequently, the same model service may face varying request pressures and performance requirements at different stages.

\begin{table}[th]
\centering
\caption{Different LLM workloads have different optimization objectives.}
\label{tab:workload}
\small
\begin{tabular}{lll}
\toprule
\textbf{Application} &
\textbf{Characteristics} &
\textbf{Primary Goal} \\
\midrule

Chat &
Interactive requests &
Low latency \\
\addlinespace
RAG &
Long-context inputs &
\makecell[l]{Prefill efficiency \\ and memory usage} \\
\addlinespace
Agent &
Multi-stage execution &
End-to-end latency \\
\addlinespace
\makecell[l]{Offline\\generation} &
Large-scale workload &
High throughput \\
\bottomrule
\end{tabular}
\end{table}

Regarding parallelism strategy adjustments, data parallelism already possesses mature runtime elasticity, allowing service capacity to be adjusted via instance scaling, sleep/wake mechanisms, etc. However, dynamically adjusting model parallelism (Tensor Parallelism/Pipeline Parallelism, TP/PP) is significantly more challenging due to its deep coupling with multiple runtime states. Specifically, changes in TP/PP topology not only affect model parameter sharding but also alter worker mapping, GPU communication patterns, and KV cache storage layouts. Therefore, changing the model parallelism configuration is not merely a parameter modification; it requires a synchronized transformation of the entire inference runtime state. As a result, mainstream LLM inference systems typically determine the model parallelism topology during service startup and keep it fixed throughout the service lifecycle. These systems construct runtime states bound to a fixed TP/PP configuration. While this design simplifies system implementation, it restricts model parallelism topology to a static deployment parameter, preventing the system from adapting execution strategies to dynamic workload changes.

Currently, when an LLM service needs to change its TP/PP configuration, the only viable approach is typically a restart-based switch: stopping the current service instance, reloading the model according to the new topology, and reinitializing the runtime environment. This approach suffers from several drawbacks. First, the service downtime is substantial, making it difficult to meet the low-latency and high-availability requirements of online LLM services. Second, existing KV cache states cannot be reused, necessitating redundant prefill computations. Finally, reloading models, creating workers, and initializing communication environments incur significant overhead, rendering dynamic model parallelism adjustment impractical for real-world deployments.

Achieving low-pause runtime TP/PP reconfiguration requires addressing three key challenges. First, we must decouple runtime states—including model weights, communication states, and worker lifecycles—from the fixed TP/PP topology to enable state transitions between different model parallelism configurations. Second, we must address KV cache migration during topology changes, as Pipeline Parallelism alters Transformer layer ownership and Tensor Parallelism changes KV head partitioning. Third, we must ensure online service continuity, minimizing service pauses while completing the transition of model, communication, and cache states.

To address these challenges, we propose \textbf{ReMP (Runtime Model-Parallelism Reconfiguration)}, a runtime model parallelism reconfiguration framework designed for LLM serving. The core insight of ReMP is to transform model parallelism topology from a fixed deployment-time configuration into a runtime-adjustable resource, enabling LLM services to dynamically adapt TP/PP configurations in response to workload variations.

ReMP first introduces a model parallelism state decoupling mechanism that disentangles model weights, communication states, worker lifecycles, and associated runtime metadata from the fixed TP/PP topology. This enables state transitions between different model parallelism configurations without requiring a full service restart. Building on this, ReMP designs a two-dimensional KV cache migration mechanism that handles layer ownership changes (due to Pipeline Parallelism) and KV head partitioning changes (due to Tensor Parallelism) separately. This allows existing KV caches to be reused under the new model parallelism configuration, avoiding cache loss and redundant prefill overhead inherent in traditional restart-based approaches. Furthermore, ReMP is integrated into the vLLM inference runtime, achieving end-to-end low-pause TP/PP topology switching through worker lifecycle management, communication state switching, and scheduling state updates.

Experimental results demonstrate that ReMP effectively addresses the high overhead of dynamic model parallelism adjustment in LLM serving. We evaluated ReMP on an eight-H100 server across multiple LLMs (ranging from 7B to 70B parameters, including Llama2-7B, DeepSeek-R1-Distill-Qwen-32B, Qwen3-30B-A3B, and Llama3-70B) and both four- and eight-GPU topologies. ReMP completes the tested switches in 1.0--7.4 seconds and is 2.5--25.8$\times$ faster than restart. Moreover, across request rates, the layout that ReMP selects at runtime achieves higher output throughput than either of the two fixed extremes, all-pipeline (TP1/PP8) and all-tensor (TP8/PP1), and in most cases lowers TTFT and TPOT as well. These results confirm that model parallelism topology can be transformed from a static deployment-time configuration into a runtime-adjustable system resource in LLM serving.

This paper makes the following contributions:
\begin{itemize}
  \item We propose ReMP, a runtime model parallelism reconfiguration framework for LLM serving.
  \item We design a model parallelism state decoupling mechanism and a two-dimensional KV cache migration mechanism to address state coupling and cache migration challenges during TP/PP reconfiguration.
  \item We implement a ReMP prototype based on vLLM and evaluate it across four LLMs and both four- and eight-GPU topologies on an eight-H100 node. Experimental results demonstrate that ReMP significantly reduces TP/PP topology switching costs, validating the feasibility of runtime model parallelism adjustment in practical LLM serving environments.
\end{itemize}

\section{Related Work}
\label{sec:related}

Recent LLM serving systems optimize scheduling, execution, and memory management under fixed deployment configurations. ORCA~\cite{orca}, vLLM~\cite{vllm}, SGLang~\cite{sglang}, FastServe~\cite{fastserve}, DeepSpeed-FastGen~\cite{deepspeed-fastgen}, LightLLM~\cite{lightllm}, and TGI~\cite{tgi} improve serving efficiency through various runtime optimizations. Another line of work explores phase-aware and disaggregated serving, including Sarathi-Serve~\cite{sarathi}, Splitwise~\cite{splitwise}, DistServe~\cite{distserve}, and Mooncake~\cite{mooncake}, which exploit prefill/decode characteristics and KV-cache management to improve resource utilization. Large-model inference systems such as FasterTransformer~\cite{fastertransformer}, TensorRT-LLM~\cite{tensorrtllm}, AlpaServe~\cite{alpaserve}, FlexGen~\cite{flexgen}, DeepSpeed Inference~\cite{deepspeed-inference}, vAttention~\cite{vattention}, and Punica~\cite{punica} further optimize distributed execution and memory efficiency. Recent adaptive serving systems, including Llumnix~\cite{llumnix}, SpotServe~\cite{spotserve}, and PipeLive~\cite{pipelive}, introduce runtime flexibility through request migration, elastic resource management, or pipeline adaptation. However, these systems generally treat the model-parallel topology as a static deployment choice or only adapt at the instance/pipeline level. In contrast, ReMP focuses on low-downtime runtime reconfiguration of both tensor and pipeline parallelism within a running LLM serving instance.

\section{System Design}
\label{sec:design}


The key idea of ReMP is to decouple runtime states that are traditionally bound to the launch-time topology. Specifically, ReMP decouples model weights from GPU shard layouts by persisting the full model state in CPU shared memory; decouples KV cache from a fixed TP/PP layout by redistributing cache tensors along both the layer dimension and the KV-head dimension; decouples communication groups from a single topology by preconstructing parallel-state snapshots; and decouples worker lifetimes from the active world size through a standby/wakeup mechanism.

\subsection{Architecture Overview}
\label{subsec:architecture}

\begin{figure*}[t]
  \centering
  \includegraphics[width=\textwidth]{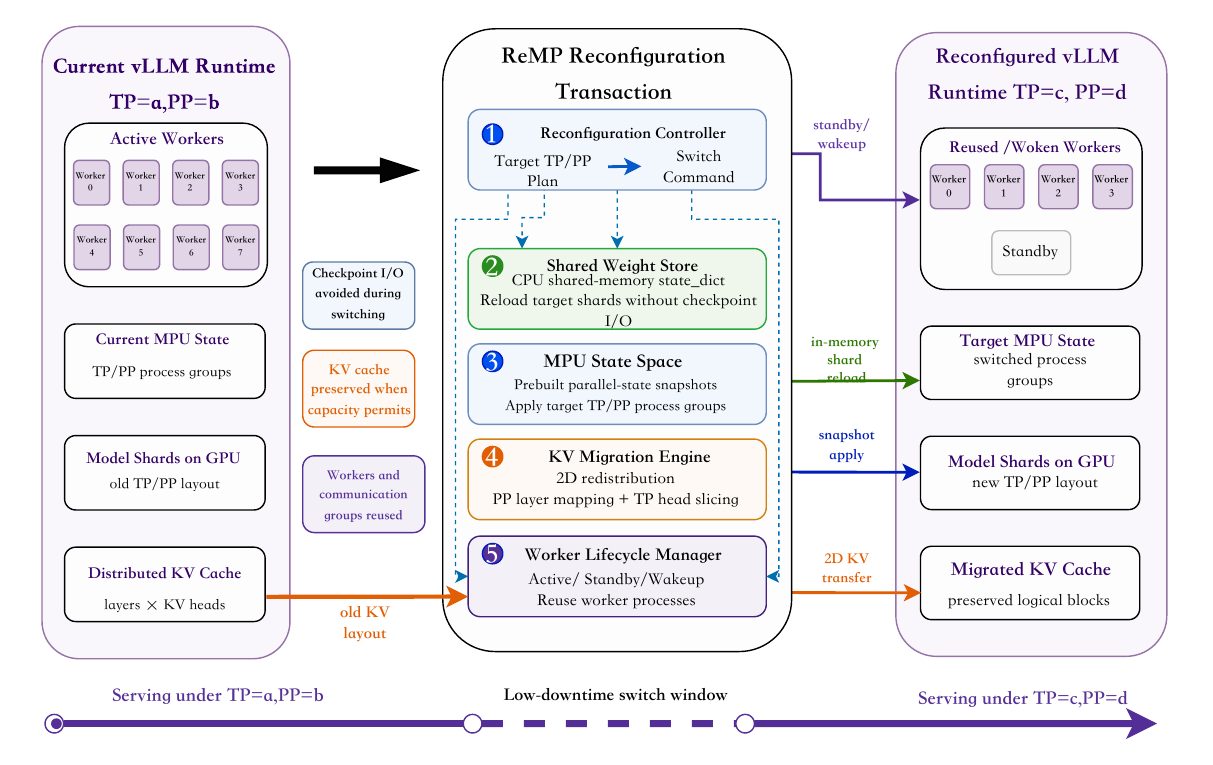}
  \caption{ReMP architecture. ReMP decouples model weights, KV cache, communication groups, and worker lifetimes from a fixed TP/PP topology, enabling low-downtime runtime reconfiguration in vLLM.}
  \Description{A block diagram of the ReMP architecture. The left side shows the serving runtime under the source topology, with active workers, the current parallel state, GPU model shards, and distributed KV cache. The middle shows the reconfiguration transaction and its six components: the Reconfiguration Controller, Shared Weight Store, MPU State Space, Worker Lifecycle Manager, KV Migration Engine, and Scheduler Adapter. The right side shows the reconfigured runtime under the target topology.}
  \label{fig:remp-overview}
\end{figure*}

Figure~\ref{fig:remp-overview} illustrates the architecture of ReMP. ReMP enables runtime TP/PP topology switching by decoupling topology-dependent runtime states, including model weights, KV cache, communication states, and worker resources. It consists of six components: the Reconfiguration Controller, Shared Weight Store, MPU State Space, Worker Lifecycle Manager, KV Migration Engine, and Scheduler Adapter.

The Reconfiguration Controller coordinates the entire switching transaction. The Shared Weight Store decouples model parameters from a fixed topology by maintaining a reusable global model state, allowing target GPU shards to be reconstructed without checkpoint reloading. The MPU State Space stores parallel-state snapshots for candidate TP/PP configurations and enables fast switching of communication states without rebuilding process groups. The Worker Lifecycle Manager maintains reusable worker resources through active and standby states, avoiding costly worker destruction and recreation.

The KV Migration Engine preserves reusable inference states by migrating KV cache according to topology changes. It handles both pipeline-parallel layer redistribution and tensor-parallel KV-head redistribution, enabling cache reuse across different TP/PP layouts. The Scheduler Adapter updates scheduling metadata and KV cache configuration after switching to ensure that the serving runtime remains consistent with the new topology.

Together, these components transform model-parallelism adjustment from a restart-based reconstruction process into an online state transformation process, enabling low-downtime runtime reconfiguration. Table~\ref{tab:state-decoupling} contrasts how each runtime state is handled by a restart-based switch and by ReMP.

\begin{table}[t]
  \centering
  \caption{ReMP decouples runtime states that are traditionally bound to the launch-time topology.}
  \label{tab:state-decoupling}
  \small
  \renewcommand{\arraystretch}{1.15}
  \scalebox{0.9}{
  \begin{tabular}{p{1.8cm}p{3.2cm}p{3.2cm}}
    \toprule
    \textbf{Runtime State} & \textbf{Restart-based Switch} & \textbf{ReMP} \\
    \midrule
    Model weights & Reload checkpoints & Reshard via CPU memory \\
    KV cache & Discard & 2D layer/head migration \\
    Comm groups & Destroy \& recreate & Switch prebuilt snapshots \\
    Workers & Terminate \& restart & Standby/wakeup reuse \\
    Scheduler & Reinitialize & Update cache config \\
    \bottomrule
  \end{tabular}
  }
\end{table}

\subsection{Reconfiguration Transaction}
\label{subsec:transaction}

ReMP treats each TP/PP topology switch as a controlled reconfiguration transaction. Given a source topology $T_{old}$ and a target topology $T_{new}$, ReMP transforms topology-dependent runtime states while keeping the serving instance alive.

A switch proceeds through four phases: pausing the scheduler, preparing target workers and MPU state, reconstructing model shards and migrating KV cache, and updating scheduler metadata before resuming execution. The scheduler resumes only after model shards, KV cache, communication states, and scheduling states become consistent with $T_{new}$.

ReMP reduces switching downtime by overlapping model reconstruction and KV migration. Model reconstruction reads target parameter slices from the shared weight store, while KV migration transfers live cache states according to the new topology. Since these operations modify independent runtime states, they can execute concurrently after the target runtime structure is prepared, reducing the critical path to the slower of the two operations.

The switching procedure handles three cases depending on the change in world size. If $TP_{old} \times PP_{old} = TP_{new} \times PP_{new}$, the active worker set remains unchanged; ReMP only needs to migrate KV cache, apply the target MPU state, and reload model shards. If the target world size is smaller, some old workers will become standby workers after the switch. Since these workers may still hold KV slices required by the target topology, ReMP migrates their useful KV state before removing them from the active worker set. If the target world size is larger, ReMP wakes up standby workers, synchronizes the message-queue ring index so that they can receive control and KV-transfer messages, and then adds them to the target topology.


\subsection{Shared Model Weight Store}
\label{subsec:shared-weight}

In conventional vLLM serving, model weights are tightly coupled with the launch-time TP/PP topology. Each worker loads and materializes its GPU shard according to its pipeline and tensor-parallel rank. Therefore, changing the topology requires restarting the service and reloading checkpoints, introducing significant loading and initialization overhead.

ReMP introduces the Shared Weight Store to decouple model parameters from a fixed topology. During service initialization, ReMP loads the full model state into CPU shared memory, which is shared across worker processes. During reconfiguration, workers reconstruct target GPU shards directly from the shared state instead of reloading checkpoint files.

For a target topology, PP determines the layer range assigned to each worker, while TP determines tensor slicing for sharded parameters such as attention and MLP weights. Each worker selects the required parameters from the shared state and materializes them according to its target rank. Replicated parameters can be accessed directly, while partitioned parameters are sliced based on the target tensor-parallel degree.

Workers still reconstruct GPU-resident shards and transfer parameters to GPU memory, but this path avoids repeated checkpoint I/O and can execute concurrently with KV cache migration during reconfiguration.

\subsection{Two-dimensional KV Cache Migration}
\label{subsec:kv-migration}

KV cache migration is the central mechanism in ReMP. In LLM serving, KV cache layout is tightly coupled with the model-parallel topology. PP determines which transformer layers are assigned to each pipeline rank and therefore where the KV cache of a layer is stored. TP determines which KV heads are assigned to each tensor-parallel rank and therefore how the head dimension of each layer's KV cache is partitioned. When TP or PP changes, the KV cache stored on one old rank may need to be split and migrated to multiple new ranks.

\subsubsection{Logical KV Cache Model}

ReMP abstracts a logical KV cache slice as $KV[l, b, h_s:h_e]$, where $l$ denotes the layer id, $b$ denotes the cache block id, and $h_s:h_e$ denotes the KV-head range. The token dimension is represented by the internal cache-block layout, while the head dimension and block size remain unchanged for the same model.

For a topology $T=(TP,PP)$, we define two ownership functions:
\[
pp\_owner(l, PP) \rightarrow pp\_rank, \quad tp\_owner(h, TP) \rightarrow tp\_rank.
\]
Given a layer $l$ and head $h$, the global rank holding the corresponding KV slice is
\[
rank(l,h,T) = rank(pp\_owner(l,PP), tp\_owner(h,TP)).
\]
Thus, topology switching must preserve the following logical mapping:
\[
KV[l,b,h] \text{ on } rank(l,h,T_{old}) \rightarrow KV[l,b,h] \text{ on } rank(l,h,T_{new}).
\]

\subsubsection{Migration Plan Construction}

ReMP constructs the migration plan from both the receiver side and the sender side. For each new rank in the target topology, ReMP first computes the target layer range from its pipeline rank, and then computes the target KV-head range from its tensor-parallel rank. It then intersects the target head range with the head ranges owned by old TP ranks. Each non-empty intersection corresponds to a KV slice that must be received from an old rank.

Each receive item can be represented as
\[
  RecvItem = (src, dst, l, B, H_{src \cap dst}),
\]
where $src$ is the old rank, $dst$ is the new rank, $l$ is the layer id, $B$ is the set of cache blocks to migrate, and $H_{src \cap dst}$ is the intersection between the source and target head ranges.

ReMP also constructs the send plan for each old rank. The send plan and receive blueprint are dual: if a new rank needs to receive a layer/head slice from an old rank, then the old rank generates a corresponding send item to that new rank.

Algorithm~\ref{alg:kv-plan} shows a simplified version of the two-dimensional migration-plan construction.

\begin{algorithm}[t]
\caption{Build 2D KV Cache Migration Plan}
\label{alg:kv-plan}
\begin{algorithmic}[1]
\Require Old topology $T_{old}$, new topology $T_{new}$, live layers $L$, live blocks $B$, number of KV heads $H$
\Ensure Send plan and receive blueprint
\ForAll{$l \in L$}
  \State $old\_pp \gets pp\_owner(l, PP_{old})$
  \State $new\_pp \gets pp\_owner(l, PP_{new})$
  \For{$ntp = 0$ to $TP_{new}-1$}
    \State $H_{target} \gets head\_range(ntp, TP_{new}, H)$
    \For{$otp = 0$ to $TP_{old}-1$}
      \State $H_{source} \gets head\_range(otp, TP_{old}, H)$
      \State $H_{overlap} \gets H_{target} \cap H_{source}$
      \If{$H_{overlap} \neq \emptyset$}
        \State $src \gets rank(old\_pp, otp, T_{old})$
        \State $dst \gets rank(new\_pp, ntp, T_{new})$
        \State $send\_plan[src] \mathbin{+}= (dst,l,B,H_{overlap})$
        \State $recv\_plan[dst] \mathbin{+}= (src,l,B,H_{overlap})$
      \EndIf
    \EndFor
  \EndFor
\EndFor
\State \Return $send\_plan, recv\_plan$
\end{algorithmic}
\end{algorithm}

\subsubsection{Local Copy and Remote Transfer}

For each migration item, if $src=dst$, the KV slice remains on the same worker after topology switching. ReMP handles this case with a local copy or view assignment, avoiding communication. If $src \neq dst$, ReMP transfers the slice from the old rank to the new rank using P2P communication. Multiple transfer items within the same layer can be batched using asynchronous send and receive operations to reduce synchronization overhead.

This design makes the actual communication volume proportional to the live KV data whose ownership changes, rather than to the full KV cache size or the number of model parameters. For topology pairs with substantial ownership overlap, more data can be handled by local copy, reducing cross-rank transfer volume.

\subsubsection{Layer-wise Streaming Migration}

A naive design would allocate a full target KV cache before releasing the old KV cache, causing the peak memory footprint to approach the sum of the old and new cache sizes. This can easily lead to out-of-memory failures for long-context workloads or high cache occupancy.

ReMP instead performs layer-wise streaming migration. For each layer to be migrated, ReMP allocates the target layer's cache storage, issues local copies and remote P2P transfers according to the migration plan, binds the target storage to the corresponding attention layer after all transfers complete, and releases the old layer storage. Since the migration working set is limited to one or a small number of layers, the peak memory overhead is much lower than materializing the entire target cache at once.

\begin{figure*}[t]
  \centering
  \includegraphics[width=\textwidth]{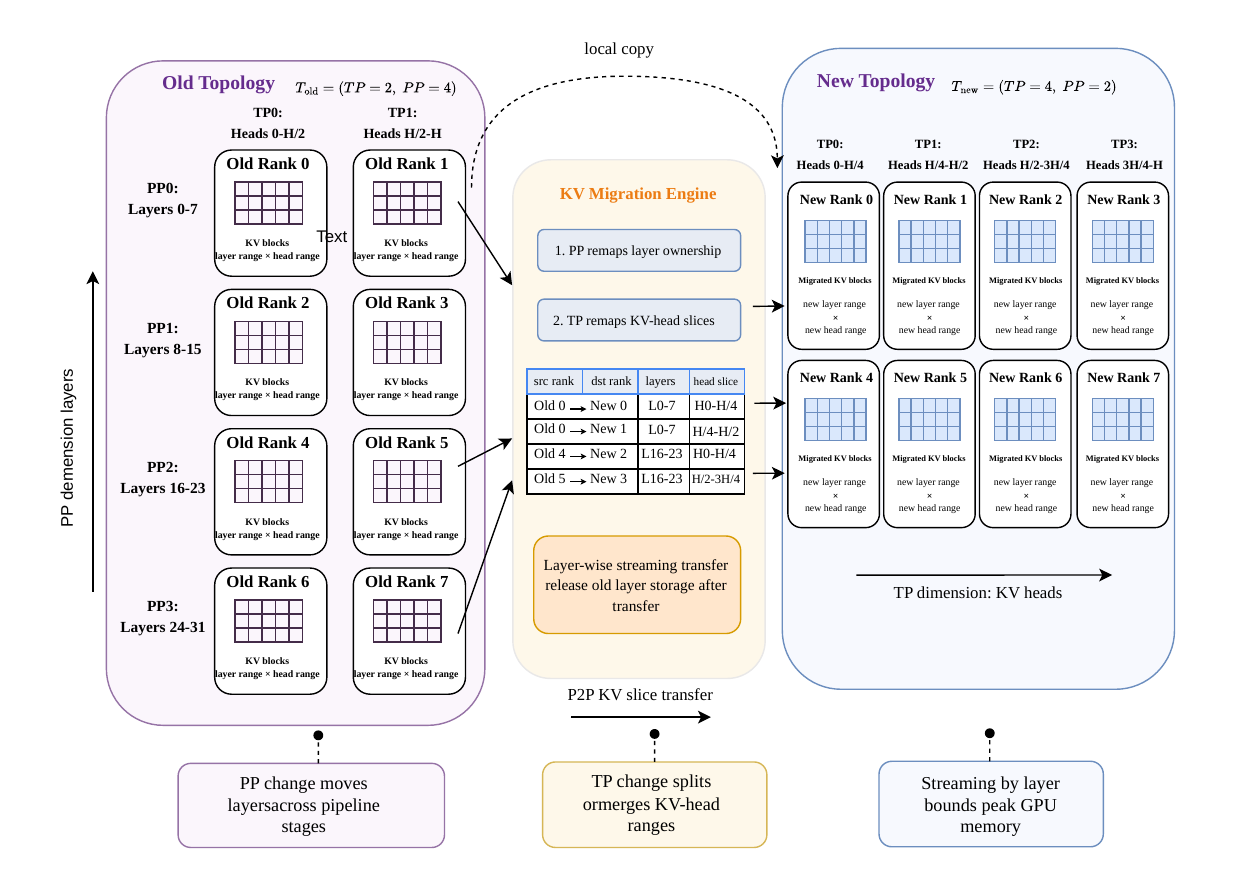}
  \caption{Two-dimensional KV cache migration. PP changes layer ownership, while TP changes KV-head slice ownership. ReMP redistributes live KV cache along both dimensions.}
  \Description{A diagram of two-dimensional KV cache migration. The left side shows the source topology, in which each rank owns a subset of transformer layers and a subset of KV heads. The right side shows the target topology after the layer and head ranges have been redistributed. The middle shows the KV Migration Engine, with arrows distinguishing cross-rank peer-to-peer transfers from local copies.}
  \label{fig:kv-migration}
\end{figure*}

Figure~\ref{fig:kv-migration} illustrates the two-dimensional KV cache migration process. PP changes remap layer ownership, while TP changes remap head-slice ownership. ReMP computes the layer/head slices required by each new rank and retrieves them from the corresponding old ranks.

\subsubsection{Correctness Invariants}

ReMP maintains the following invariants during KV cache migration.

\textbf{Layer coverage.}
For every layer assigned to a target PP rank, all live cache blocks that should be preserved for that layer must exist in the corresponding target cache storage after migration.

\textbf{Head coverage.}
For each migrated $(layer, block)$ pair, the union of head ranges owned by all target TP ranks must equal the full KV-head range, without overlapping ownership unless replication is required by the model semantics.

\textbf{Logical block identity preservation.}
The mapping maintained by the scheduler from requests, prefixes, and logical block ids to KV contents must remain valid after migration, or be updated through a deterministic block remapping, so that migrated KV cache can continue to be referenced by the scheduler.

\textbf{Capacity constraint.}
If the target topology provides less KV cache capacity than the current live cache footprint, ReMP cannot preserve all KV blocks. In this case, the scheduler preempts selected running requests, reclaims their cache blocks, and moves those requests back to the waiting queue for later recomputation.

\subsection{MPU State Space}
\label{subsec:mpu-space}

TP/PP topology determines not only model and KV cache placement, but also the communication structure. In conventional systems, TP groups, PP groups, world groups, rank mappings, and other parallel-state metadata are created at service startup and assumed to remain unchanged throughout the service lifetime. Destroying and recreating NCCL process groups during reconfiguration would add latency and introduce complex synchronization and failure-handling issues.

ReMP introduces the MPU State Space to manage communication states for multiple candidate topologies. For each candidate topology $T_i=(TP_i,PP_i)$, ReMP preconstructs a corresponding parallel-state snapshot: $MPUSpace = \{T_i \rightarrow Snapshot_i\}$.
Each snapshot contains TP groups, PP groups, rank mapping, tensor model-parallel rank, pipeline model-parallel rank, and other metadata required by vLLM's parallel state. During switching, a worker looks up the snapshot for the target topology and applies it to the global parallel state.

This design moves process-group construction off the switching critical path and into the initialization phase. The trade-off is that the candidate topology set must be bounded and known in advance. Supporting arbitrary topologies would require a dynamic group cache or on-demand group construction. 

\subsection{Worker Lifecycle Management}
\label{subsec:worker-management}

The vLLM multiprocess executor normally creates a fixed set of workers according to the launch-time world size. If the world size changes through a restart-based path, the system must destroy old workers and create new ones, reinitializing the CUDA runtime, model objects, communication state, and message queues. ReMP avoids this overhead through explicit worker lifecycle management.

ReMP classifies workers into active and standby states. Active workers participate in model execution, communication, and KV cache management under the current topology. Standby workers retain process resources but do not execute requests. When the target world size is smaller than the current world size, ReMP removes extra workers from the active set and places them into standby. When the target world size is larger, ReMP wakes standby workers and adds them back to the active set.

During scale-down, ReMP must migrate KV cache before moving extra workers to standby. This is because a worker that will leave the active set may still hold KV slices required by the target topology. If such a worker were put into standby before migration, its KV state might become inaccessible to the target topology. During scale-up, ReMP first wakes standby workers and synchronizes the message-queue ring index so that the new workers can correctly receive executor messages and KV-transfer requests. The workers then apply the target MPU state, load their target model shards from the shared weight store, and receive migrated KV cache.

A standby worker retains a lightweight runtime context and the handles needed for fast wakeup. ReMP can therefore wake the target workers and reconfigure their communication state without repeating process creation, runtime initialization, and device-context setup, removing process-management overhead from the switching path.

\subsection{Scheduler and KV Block Adaptation}
\label{subsec:scheduler-adapter}

Topology switching changes the KV cache layout and available cache capacity on each worker. ReMP therefore updates scheduler state to match the target topology after migration.

During switching, the scheduler temporarily pauses scheduling decisions and freezes KV metadata to provide a consistent view for cache migration. After switching, the Scheduler Adapter regenerates the KV cache configuration, including layer placement, cache capacity, tensor shapes, and attention-layer bindings, and updates the block manager accordingly.

If the target topology provides additional KV capacity, the scheduler expands the available cache blocks. If the capacity decreases, ReMP first reclaims unused blocks and, when necessary, preempts selected requests to release cache space. This ensures that the scheduler continues operating correctly under the new topology.

In addition, PP changes modify the pipeline-stage structure. ReMP refreshes pipeline batch metadata after switching so that subsequent scheduling decisions are consistent with the target execution layout.

The scheduler resumes only once the target worker set, MPU state, migrated KV cache, model shards, and cache configuration are all in place, so no request observes a partially migrated shard. Until then the source topology stays intact, and an early failure falls back to it.

\section{Evaluation}
\label{sec:evaluation}




Our evaluation answers two questions. \emph{How expensive is a runtime TP/PP switch?} Section~\ref{sec:eval-reconfiguration} quantifies ReMP's switching cost against restart-based reconfiguration and identifies which operations dominate that cost. \emph{What does a cheap switch buy?} Section~\ref{sec:eval-serving-performance} shows that different models and request rates favor different TP/PP configurations, and compares the layout that ReMP selects at runtime against the two fixed extremes of the candidate space, all-pipeline and all-tensor.
\subsection{Experimental Setup}
\label{subsec:eval-setup}
  \begin{table*}[t]
  \centering
  \caption{Model configurations used in the evaluation.}
  \label{tab:eval-models}
  \small
  \begin{tabular}{lccccccc}
    \toprule
    \textbf{Model} &
    \textbf{Arch.} &
    \textbf{Params} &
    \textbf{Active Params} &
    \textbf{Layers} &
    \textbf{Hidden Size} &
    \textbf{Attn. Heads} &
    \textbf{KV Heads} \\
    \midrule
    Llama2-7B &
    Dense &
    7B &
    7B &
    32 &
    4096 &
    32 &
    32 \\

    Llama3-70B &
    Dense &
    70B &
    70B &
    80 &
    8192 &
    64 &
    8 \\

    DeepSeek-R1-Distill-Qwen-32B &
    Dense &
    32B &
    32B &
    64 &
    5120 &
    40 &
    8 \\

    Qwen3-30B-A3B &
    MoE &
    30.5B &
    3.3B &
    48 &
    2048 &
    32 &
    4 \\
    \bottomrule
  \end{tabular}
\end{table*}

\textbf{Hardware.}
We evaluate ReMP on a server equipped with 8 NVIDIA H100 GPUs, an AMD EPYC 7R13 CPU, and 2TB host memory.

\textbf{Software.}
We implement ReMP on vLLM V1 with runtime TP/PP switching, shared-memory model reuse, two-dimensional KV cache migration, prebuilt communication snapshots, and worker management. The two systems use the same software stack. The baseline alone restarts the service to change TP/PP topology.

\textbf{Models.}
Table~\ref{tab:eval-models} summarizes the models used in our evaluation. We use four representative LLMs ranging from 7B to 70B parameters, including both dense and MoE architectures. For brevity, we abbreviate DeepSeek-R1-Distill-Qwen-32B as DeepSeek-32B in the remainder of this section. These models differ in parameter size, number of layers, hidden dimension, attention layout, and KV-head configuration, which affect both model shard reconstruction and KV cache migration during TP/PP reconfiguration.

\textbf{Workloads.}
For serving-performance experiments, we use BurstGPT-derived request traces and replay them at different request rates. Each run uses the same request sequence across TP/PP configurations to ensure that performance differences come from the serving topology rather than workload variation.

\textbf{Baselines.}
For reconfiguration-cost experiments, we compare ReMP with a restart-based baseline. The restart baseline terminates the current serving instance and launches a new one under the target TP/PP topology, reloading model checkpoints, reconstructing runtime state, and discarding existing KV cache.

For serving-performance experiments, we compare ReMP with fixed TP/PP baselines, each of which holds one static configuration for the whole workload. ReMP instead uses low-downtime switching to select a configuration from measured performance, as described in Section~\ref{sec:eval-serving-performance}; the figures label the resulting curve \textbf{ReMP}.

\textbf{Metrics.}
For reconfiguration cost, we report restart-based switching time, ReMP switching time, speedup over restart, and ReMP's internal switching-time breakdown. The breakdown includes model loading from the shared weight store and KV cache transfer. For serving performance, we report mean TTFT, mean TPOT, and output throughput, where \emph{output throughput} is the number of generated tokens per second sustained by the serving instance. Output throughput is also the ranking metric: preselection retains the candidate layout that achieves the highest output throughput.
\subsection{Reconfiguration Cost}
\label{sec:eval-reconfiguration}

\begin{figure*}[t]
  \centering
  \includegraphics[width=\textwidth]{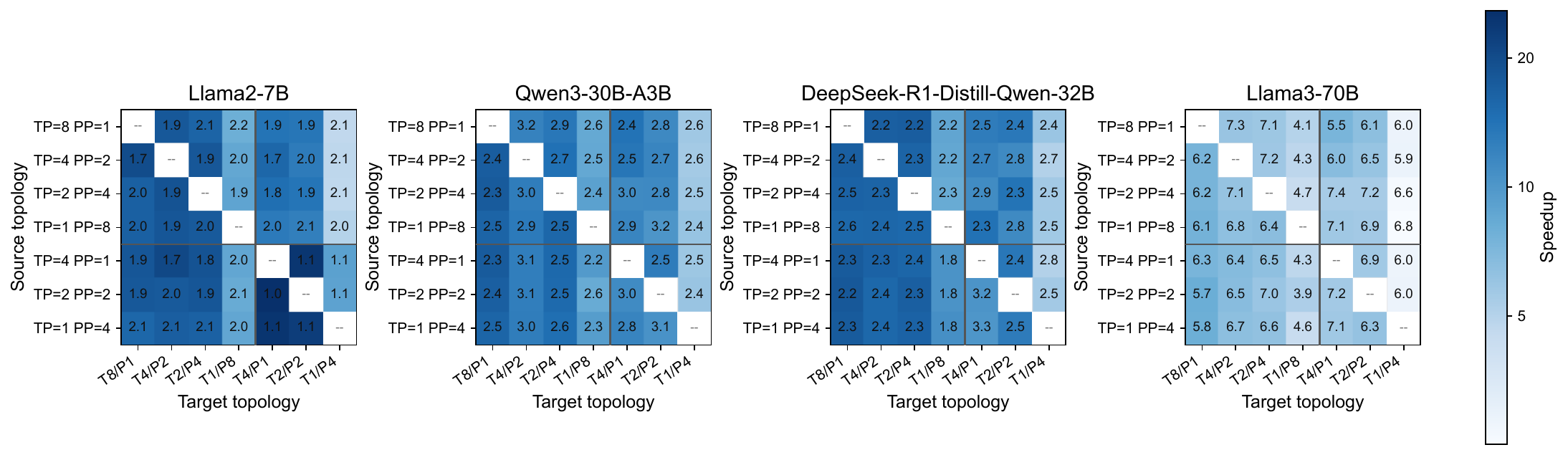}
  \caption{End-to-end ReMP switching time and speedup over restart on H100. Rows and columns contain the same four eight-GPU and three four-GPU topologies. Horizontal and vertical separators divide eight- and four-GPU layouts, exposing eight-to-eight, four-to-eight, eight-to-four, and four-to-four transitions. Cell text reports $T_{\text{ReMP}}$ in seconds, while color (on a logarithmic scale) reports $T_{\text{restart}}/T_{\text{ReMP}}$. Diagonal cells are omitted because the topology is unchanged.}
  \Description{Four seven-by-seven heatmaps compare ReMP with restart for Llama2-7B, Qwen3-30B-A3B, DeepSeek-32B, and Llama3-70B across four- and eight-GPU topology changes. Across all 168 non-identity switches, ReMP takes 1.0 to 7.4 seconds and achieves speedups from 2.5 to 25.8 times.}
  \label{fig:e2e-switching}
\end{figure*}

We compare ReMP with restart-based reconfiguration on H100 across four models using the symmetric $7\times7$ source--target matrix in Figure~\ref{fig:e2e-switching}. Its rows and columns contain the same four eight-GPU layouts (TP8/PP1, TP4/PP2, TP2/PP4, and TP1/PP8) followed by three four-GPU layouts (TP4/PP1, TP2/PP2, and TP1/PP4). The four matrix regions therefore capture eight-to-eight reconfiguration, four-to-eight scale-up that wakes standby workers, eight-to-four scale-down that deactivates excess workers, and four-to-four reconfiguration among TP4/PP1, TP2/PP2, and TP1/PP4. The restart baseline rebuilds the target service at its corresponding GPU count, whereas ReMP transforms topology-dependent state within the running service. All four models have complete non-identity measurements in all four regions.

Figure~\ref{fig:e2e-switching} contains all 168 non-identity source--target comparisons: 48 each for eight-to-eight, four-to-eight, and eight-to-four transitions, plus 24 four-to-four transitions. ReMP switches Llama2-7B in 1.0--2.2 seconds, corresponding to a 4.7--25.8$\times$ speedup over restart. Qwen3-30B-A3B requires 2.2--3.2 seconds and achieves a 5.6--17.4$\times$ speedup. DeepSeek-32B requires 1.8--3.3 seconds and achieves a 5.1--17.9$\times$ speedup. Llama3-70B requires 3.9--7.4 seconds and remains 2.5--8.0$\times$ faster than restart over its 42 transitions. Its six four-to-four switches take 6.0--7.2 seconds and are 2.8--6.2$\times$ faster than restart. Across the complete matrix, speedup ranges from 2.5$\times$ to 25.8$\times$, with a median of 11.3$\times$.

The two scale directions expose an asymmetry that is hidden by an eight-GPU-only comparison. Eight-to-eight and four-to-eight transitions achieve median speedups of 13.3$\times$ and 13.7$\times$, respectively. Eight-to-four scale-down has a lower median of 8.8$\times$ because restarting some four-GPU targets, especially TP1/PP4, is faster than restarting an eight-GPU target; nevertheless, ReMP remains faster in every scale-down cell. The 24 four-to-four switches have a median speedup of 9.6$\times$ and include the overall maximum of 25.8$\times$.

The cell-to-cell variation reflects how much topology-dependent state must be transformed: PP changes remap transformer layers across pipeline stages, and TP changes redistribute tensor slices and KV-head ownership across tensor-parallel ranks. Larger models and transitions requiring more extensive shard reconstruction therefore cost more, but all tested transitions remain at the second scale.

These speedups reflect two major costs that ReMP removes from the switching critical path. First, ReMP keeps workers, device contexts, and candidate communication groups alive across switches. It therefore avoids process creation, worker initialization, CUDA-runtime setup, and communication-group construction during reconfiguration. Second, ReMP loads topology-specific model shards from a dedicated CPU shared-memory weight store. A restart can benefit from a warm operating-system page cache, but it still traverses the filesystem and framework checkpoint-loading path before rebuilding GPU-resident shards. Direct shared-memory access can remove much of this indirection and provide faster, less variable model loading than relying on the page cache alone. Together, these mechanisms make topology changes an in-memory state transformation rather than a full service reconstruction.

\begin{figure}[t]
  \centering
  \includegraphics[width=\linewidth]{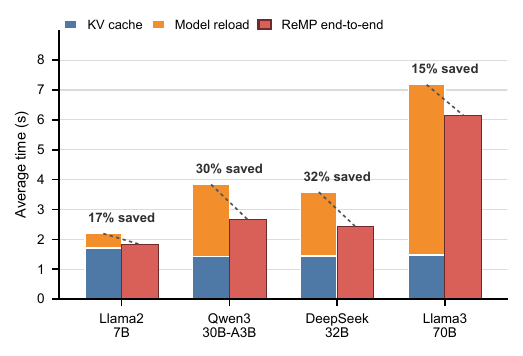}
  \caption{Effect of overlapping model-shard reloading with KV cache construction or migration on H100. Stacked bars show the sequential reference $T_{\text{model}}+T_{\text{kv}}$; red bars show measured end-to-end ReMP time with overlap. Annotations report the reduction for each model.}
  \Description{A grouped bar chart for four models on H100. For each model, a stacked bar shows the sequential sum of KV cache and model reloading time, and a red bar shows the lower measured end-to-end ReMP switching time with overlap. The reductions are 17, 30, 32, and 14 percent from 7B to 70B.}
  \label{fig:overlap-breakdown}
\end{figure}

Figure~\ref{fig:overlap-breakdown} averages the non-identity transitions for each model and shows that overlapping model reloading with KV cache processing reduces their sequential sum by 14--32\%. The reduction is largest for Qwen3-30B-A3B and DeepSeek-32B (30\% and 32\%), whose two components are comparably sized, and smaller for Llama2-7B and Llama3-70B (17\% and 14\%) when one component dominates. Thus, the overlapped critical path is bounded by the slower component.

  \begin{figure*}[t]
  \centering
  \includegraphics[width=\textwidth]{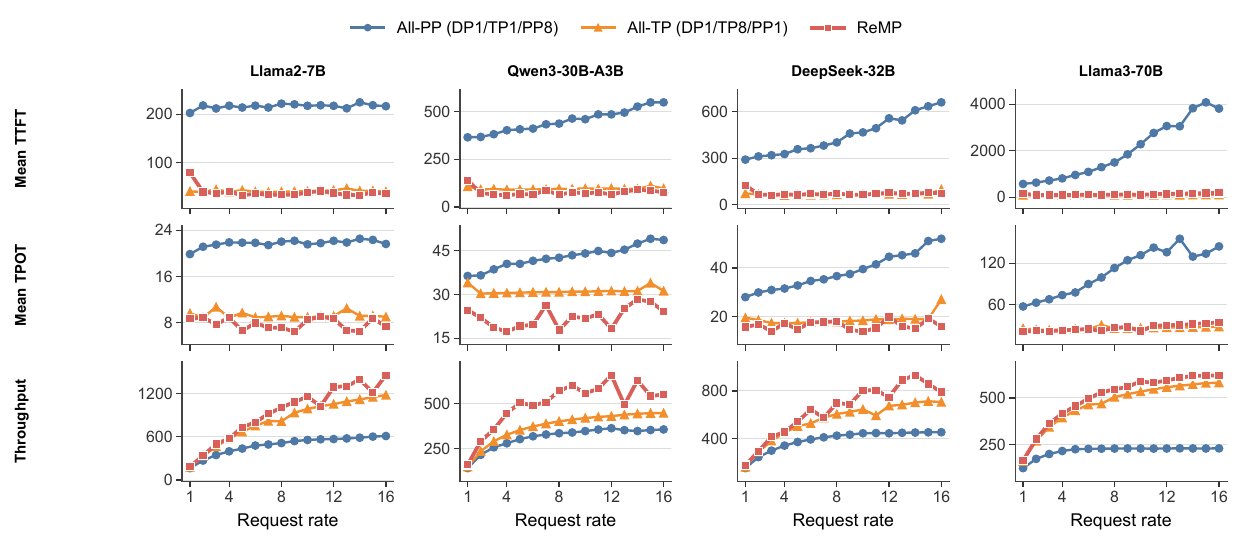}
  \caption{Serving performance on an 8$\times$H100 node, replaying BurstGPT at integer request rates 1--16. All-PP and All-TP are the fixed DP1/TP1/PP8 and DP1/TP8/PP1 baselines. At each model--rate pair, ReMP retains the layout with the highest output throughput. The plotted mean TTFT, mean TPOT, and output throughput are measured on that layout, and every plotted case completes without failure.}
  \label{fig:exp2_h100}
  \Description{A three-row by four-column line chart comparing mean TTFT, mean TPOT, and output throughput for fixed all-pipeline, fixed all-tensor, and ReMP-selected parallel layouts across four models and sixteen request rates on eight H100 GPUs.}
\end{figure*}

\subsection{Serving Performance under Different Request Pressure}
\label{sec:eval-serving-performance}

Section~\ref{sec:eval-reconfiguration} shows that ReMP switches model-parallel topologies at second scale. We now evaluate what that capability buys a serving system. Because a switch is cheap, ReMP can measure candidate layouts rather than predict them: it exercises the candidate DP/TP/PP layouts on an initial window of requests and serves the remaining requests with the layout that measured best. We refer to this as \emph{preselection}, and we ask how the layout it retains compares with the fixed extremes of the same candidate space across models and request rates.

\subsubsection{Setup}
On a single 8$\times$H100 node, we replay the BurstGPT dataset at integer request rates (RPS) from 1 to 16 against Llama2-7B, Qwen3-30B-A3B, DeepSeek-32B, and Llama3-70B, under seven layouts satisfying $\mathrm{DP}\times\mathrm{TP}\times\mathrm{PP}=8$.

Figure~\ref{fig:exp2_h100} compares ReMP with two fixed extremes at each model--rate pair. All-PP (DP1/TP1/PP8) maximizes pipeline depth, whereas All-TP (DP1/TP8/PP1) devotes all eight GPUs to tensor parallelism. Preselection ranks candidates by output throughput, and the plotted mean TTFT, mean TPOT, and output throughput all come from the single layout that this ranking retains.

\subsubsection{Preselection and Elastic Data Parallelism}
ReMP initializes two data-parallel replicas before preselection, so the state required by DP2 is already resident and the DP degree becomes a routing decision rather than a construction step. If preselection retains a DP1 layout, ReMP places one replica in vLLM sleep mode, which releases most of its GPU memory, and routes all requests to the active replica; if it retains a DP2 layout, both replicas stay awake and share the request stream. Waking a sleeping replica takes seconds, so the two states remain interchangeable at runtime. 

\subsubsection{Topology Selection}
Across 64 model--rate points, preselection retains DP2 at 15 and DP1 at 49; no layout or DP degree wins throughout the sweep. Table~\ref{tab:selection-summary} summarizes the model-specific selections.

\begin{table*}[t]
  \centering
  \caption{Topology-selection frequency and observed peak output throughput.}
  \label{tab:selection-summary}
  \begin{tabular}{lccccc}
    \toprule
    \textbf{Model} & \textbf{Most frequent layout} & \textbf{Selections} & \textbf{DP2 points} & \textbf{Peak RPS} & \textbf{Peak output tok/s} \\
    \midrule
    Llama2-7B & DP2/TP2/PP2 & 10/16 & 10/16 & 16 & 5,882.6 \\
    Qwen3-30B-A3B & DP1/TP4/PP2 & 16/16 & 0/16 & 12 & 2,537.5 \\
    DeepSeek-32B & DP1/TP4/PP2 & 14/16 & 1/16 & 14 & 3,602.9 \\
    Llama3-70B & DP1/TP4/PP2 & 12/16 & 4/16 & 16 & 2,400.5 \\
    \bottomrule
  \end{tabular}
\end{table*}

Selections are model dependent. Qwen3-30B-A3B consistently favors TP4/PP2. DeepSeek-32B and Llama3-70B occasionally switch TP or DP degree, while Llama2-7B favors DP2/TP2/PP2, using its smaller footprint for replica-level concurrency. Thus, DP helps only when this concurrency outweighs the loss of TP/PP resources per replica.

\subsubsection{Throughput Scaling}
ReMP improves on both fixed extremes across the sweep. Averaged over the 16 rates, its per-rate improvement in output throughput over All-PP is 85.2\%, 56.0\%, 60.8\%, and 126.7\% for Llama2-7B, Qwen3-30B-A3B, DeepSeek-32B, and Llama3-70B, respectively; over All-TP it is 12.0\%, 31.4\%, 15.1\%, and 6.9\%. The gap to All-PP widens with load: at RPS~16 the improvements reach 138.4\%, 54.1\%, 73.9\%, and 168.9\%, as the deep fixed pipeline becomes the binding constraint on throughput.

From RPS~1--4 to RPS~13--16, ReMP's average gain over All-PP rises from 31.5\% to 125.7\% for Llama2-7B, from 26.8\% to 91.8\% for DeepSeek-32B, and from 66.3\% to 166.9\% for Llama3-70B; Qwen3-30B-A3B gains 36.3\% at low load and 56.9\% at high load. Pipeline depth therefore becomes increasingly restrictive as more requests compete for stage capacity.

\subsubsection{Latency and Objective Trade-offs}
Latency follows the same pattern. At RPS~16, All-PP's mean TTFT reaches 216.7, 547.6, 661.1, and 3814.9~ms for the four models, whereas ReMP records 35.2, 77.4, 72.8, and 192.6~ms---reductions of 6.2$\times$, 7.1$\times$, 9.1$\times$, and 19.8$\times$---and All-PP's mean TPOT is 2.0--4.3$\times$ higher at the same rate. Against All-TP, ReMP raises output throughput at RPS~16 by 23.4\%, 23.0\%, 11.9\%, and 7.0\%, and for the first three models it does so at lower mean TTFT and TPOT as well. Llama3-70B is the one case where the two goals conflict: the retained single-replica TP4/PP2 layout accepts 73.1\% higher mean TTFT and 23.8\% higher mean TPOT than All-TP to obtain that 7.0\%. Preselection optimizes throughput by construction, and a latency-constrained objective would have retained DP2/TP4/PP1 instead, which at this rate is the lowest-latency of the seven candidates.

The full sweep provides a broader view than the RPS~16 endpoint. Relative to All-PP, ReMP lowers both mean TTFT and mean TPOT at all 64 model--rate points. Relative to All-TP, it lowers mean TTFT at 34 points and mean TPOT at 49 points. The remaining cases reflect an objective trade-off rather than a selection failure: preselection maximizes output throughput and does not require every latency metric to improve simultaneously.

This distinction is clearest for Llama3-70B. ReMP produces strictly higher output throughput than All-TP at all 16 rates, but it lowers mean TTFT at only one and mean TPOT at eight. For Qwen3-30B-A3B, the selected TP4/PP2 layout improves output throughput at all 16 rates and lowers mean TPOT at all 16. These contrasting patterns show why a serving policy should expose its objective explicitly. A latency-constrained policy can use the same switching mechanism while ranking candidate layouts under a different utility function.

\subsubsection{Implications}
TP/PP reconfiguration changes per-request efficiency, whereas DP changes replica-level concurrency. Because the best combination depends on model and request rate, second-scale switching lets ReMP measure rather than commit to a launch-time layout. Output throughput ranks layouts only within the same model and request stream; cross-model token rates are not compared.

\section{Conclusion}

This paper presents ReMP, which decouples model weights, KV cache, communication state, workers, and scheduler metadata from a fixed TP/PP topology to reconfigure model parallelism without a full service restart.

Our evaluation shows that ReMP completes the tested topology switches in 1.0--7.4 seconds and is 2.5--25.8$\times$ faster than restart-based switching across four- and eight-GPU topologies. Overlapping model reloading with KV cache processing reduces the complete switching path by 14--32\%. Under dynamic request pressure, the layout selected at runtime improves output throughput over both fixed extremes across the request-rate sweep.

Together, these results show that ReMP can jointly adapt per-request execution efficiency and replica-level concurrency while preserving model and KV cache state, turning model-parallel topology into a runtime-adjustable serving resource.


\end{document}